# Dynamics and Control of Infections on Social Networks

Brian G. Williams[1] and Christopher Dye[2]

1. Wits Reproductive Health and HIV Institute, University of the Witwatersrand, Johannesburg, South Africa and South African Centre for Epidemiological Modelling and Analysis, University of Stellenbosch, South Africa (Correspondence to BrianGerardWilliams@gmail.com)
2. World Health Organization, Geneva, Switzerland (dyec@who.int)

## Abstract

Random mixing in host populations has been a convenient simplifying assumption in the study of epidemics, but neglects important differences in contact rates within and between population groups. For HIV/AIDS, the assumption of random mixing is inappropriate for epidemics that are concentrated in groups of people at high risk, including female sex workers (FSW) and their male clients (MCF), injection drug users (IDU) and men who have sex with men (MSM). To find out who transmits infection to whom and how that affects the spread and containment of infection remains a major empirical challenge in the epidemiology of HIV/AIDS. Here we develop a technique, based on the routine sampling of infection in linked population groups, which shows how an Asian HIV/AIDS epidemic began in FSW, was propagated mainly by IDU, and ultimately generated most cases among the female partners of MCF (FPM). Calculation of the case reproduction numbers within and between groups, and for the whole network, provides insights into control that cannot be deduced simply from observations on the prevalence of infection. Specifically, the *per capita* rate of HIV transmission was highest from FSW to MCF, and most HIV infections occurred in FPM, but the number of infections in the whole network is best reduced by interrupting transmission to and from IDU. This network analysis can be used to guide HIV/AIDS interventions based on needle exchange, condom distribution and antiretroviral therapy. The method requires only routine data and could be applied to infections in other populations.

## 1. Introduction

Epidemiological theory assumes that infections are transmitted through random contacts between infected and uninfected people. The reality is usually different, and simple assumptions can give misleading results. One example is the spread of HIV/AIDS in 'concentrated epidemics', where populations contain small groups of people at high risk and large groups of people at low risk. Here we show that, when investigating the control of such epidemics, routinely collected data are a rich source of information. Using surveillance data to characterize the transmission network for HIV/AIDS in Vietnam we find that the best way to minimize infections in the whole population is first by targeting high-risk injection drug users, then men who have sex with men, and finally female sex workers.

Generalized epidemics of HIV/AIDS, such as those prevailing in East and southern Africa, are driven mainly by heterosexual transmission in the population at large.[1,2] Concentrated epidemics, on the other hand, are focused on networked groups of people who acquire and transmit virus by a mix of sexual transmission (between men and women and among men) and by needle injection of contaminated blood. Investigations of the structure of these networks have usually been carried out with social surveys[3,4] or by identifying transmission links with genetic markers,[5-8] in order to track the spread of infection through populations. However, the accurate reconstruction of transmission pathways by these methods is labour intensive both in the field and in the laboratory. In this paper we develop an alternative method of constructing an epidemic network based on the routine sampling, through time, of infection in linked population groups. We have used the method to gain insights into the way an epidemic of HIV/AIDS unfolded in Vietnam, and to investigate how the spread of infection can most effectively be reversed.

The control of HIV in concentrated epidemics demands different interventions for different risk groups. In Thailand the '100% Condom Programme' for Female Sex Workers, combined with other interventions, has significantly reduced HIV transmission.[9] For injecting drug users a meta-analyses suggests that access to clean needles could reduce HIV transmission by 66%[10] while another meta-analysis suggests the opiate substitution therapy could reduce transmission by 54%,[11] In generalized epidemic settings early treatment has been found to reduce transmission by 96%.[12,13] While both the impact and the cost of different combinations of interventions vary, we are concerned in this paper with the population impact that can be achieved for a given reduction in the individual risk of transmission however it is bought about.

## 2. Methods and data

This analysis focuses on the spread of an HIV/AIDS epidemic in Can Tho province, Vietnam, as described by data collected as part of the annual National Sentinel Surveillance system (1994 to 2010) and from Integrated Biological and Behavioural Surveillance surveys in 2006 and 2009.[14]

In 2010 the prevalence of HIV was highest among injection drug users (IDU: 48%), then men who have sex with men (MSM: 9.5%), followed by female sex workers (FSW: 5.8%), male clients of FSW (MCF: 1.1%) and finally female partners of men in each group (FPM: 0.5%). While the prevalence of infection is lowest in FPM, this group carries the largest number of infections, making up 49% of all infected people, because they are by far the largest group among those at risk of infection.

their male clients (MCF), each have potentially self-sustaining epidemics. They are connected through MSM and FSW who are also IDU. The female partners of men who visit sex workers (FPM) and of other men are assumed to be an epidemiological dead end, and do not infect anyone else.[14] In Figure 1, the weight of the arrows indicates the expected extent of transmission. For example, each FSW may infect many MCFs but each MCF is likely to infect relatively FSWs.

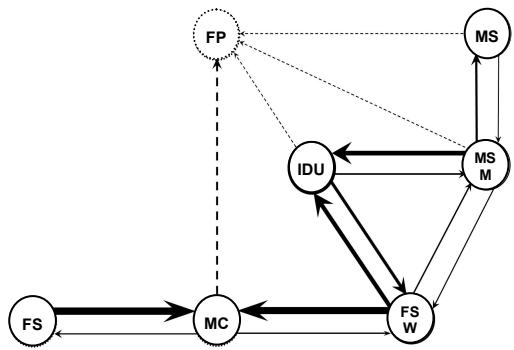

Figure 1. The network model for HIV in Can Tho Province, Viet Nam. IDU: Injection drug users; MSM: Men who have sex with men; FSW: Female sex workers; MCF: Male clients of FSWs; FPM: Female partners of MCF and other women at low risk

We use a previously constructed network including transmission within groups and all probable links between pairs of groups (Figure 1)[14]. Injection drug users (IDU), men who have sex with men (MSM), and female sex workers (FSW) and

Table 1. Risk groups, the estimated number in each group, the prevalence of infection, the number of infected people in each group, and the mean time for which people remain in each risk group as estimated for 2011. IDU: Injection drug users; MSM: Men who have sex with men; FSW: Female sex workers; MCF: Male clients of FSWs; FPM: Female partners of MCF and other women at low risk.

| Risk group | Number | Prevalence (%) | No. infected | Duration (yrs) |
|---|---|---|---|---|
| IDU | 2,716 | 49.50 | 1100 | 12 |
| MSM | 1,176 | 3.62 | 43 | 20 |
| MSM&IDU | 324 | 30.82 | 100 | 12 |
| FSW | 1,978 | 4.08 | 81 | 20 |
| FSW&IDU | 62 | 61.72 | 38 | 12 |
| MCF | 61.6k | 1.06 | 653 | 8 |
| FPM | 454k | 0.45 | 2043 | 20 |

The differential equations for the network in Figure 1, are given in the Appendix. The initial prevalence (in 1980) and the transmission parameters were varied to obtain the maximum likelihood fit to the trend data assuming binomial errors.[14] This gives the estimated size and prevalence in each group and sub-group in 2011 (Table 1) and the fitted trends shown in (Figure 2).

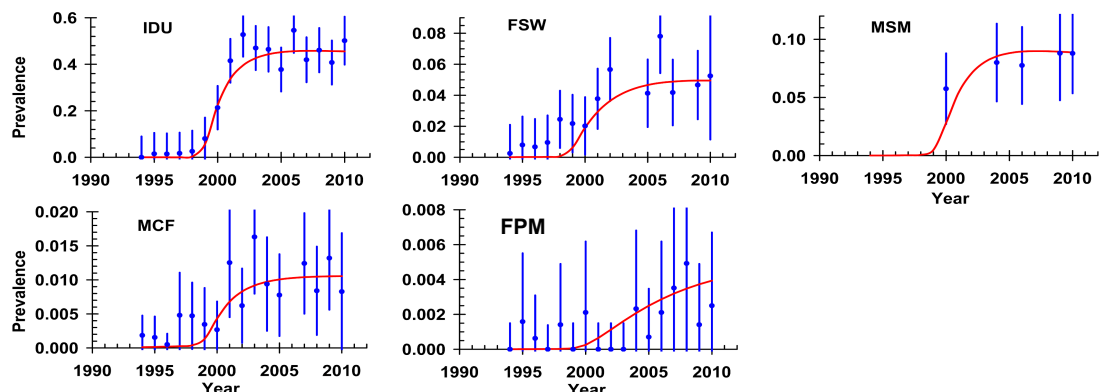

Figure 2. Trends in the prevalence of HIV over time for different risk groups in Can Tho province. IDU: Injection drug users; MSM: Men who have sex with men; FSW: Female sex workers; MCF: Male clients of FSWs; FPM: Female partners of MCF and other women at low risk.

In order to provide a quantitative guide controlling the epidemic we analyze the elements of the next-generation matrix (NGM) which give the case reproduction numbers[15] within and between groups. From the values of the coefficients in the model equations (Appendix), fitted to the time-series data (Figure 2), we obtain the elements of the NGM[15]. The principal eigenvalue of the NGM is $R_0$, the basic case reproduction number for the whole network; when $R_0 < 1$, infection will be eliminated from the network[15]. Furthermore, on the approach to elimination, the smaller the value of $R_0$, the smaller the number of people that will be infected. If a single infected case is introduced into one group then the elements of the NGM give the number of secondary cases that arise in each of the groups in the network and provide an elegantly simple method of investigating the impact of control measures, without resorting to specific numerical simulations and projections.

## 3. Results

An earlier investigation of these data[14] could not match the rapid rise in the prevalence of IDUs with the much slower rises in prevalence in other groups (Figure 2). To get a better fit to the data we assumed that infection was introduced initially among FSWs and then spread from them to IDUs. By setting the prevalence of HIV in the IDU group to zero in 1980, but allowing it to be non-zero in the other groups, we obtained the fit to the data shown in Table 1 and Figure 2, which more accurately describes the spread of infection in all to groups including IDUs.

Our first deduction from fitting the model to the time-series data is that the epidemic was probably introduced through female sex workers (FSW). From FSW it spread to injection drug users (IDU), who then became the key drivers of the epidemic. This conclusion is based on the observation that the model can accurately describe the epidemic in IDU only by assuming that HIV prevalence was zero in this group in 1980 and that infections in IDUs were introduced through the small group of FSWs who also inject drugs. The NGM gives $R_0 = 22$ for the whole network, much larger than the value of $R_0 = 4.1$ that would have been obtained by assuming random mixing among all the risk groups, assuming that they were all at equal risk, and fitting the model to time trends in the overall prevalence of HIV. The individual elements of the NGM give the values of $R_0$ for transmission within and among population groups (Table 2). In Figure 3, values of $R_0$ written in the circles apply within groups. Values of $R_0$ written on the lines connecting groups give the number of secondary cases that arise from one primary case in the source group. Values of $R_0$ written between the lines give the number of secondary cases arising in one primary group, via a linked secondary group.

Table 2. The next-generation matrix for the epidemic of HIV in Vietnam. For the whole system the eigenvalue of the dominant eigenvector gives $R_0 = 22.0$. The last row gives the dominant eigenvector. The table gives the number of secondary cases in each group in a given row, as well as for all groups combined, for one primary case in each group in a given column in an otherwise fully susceptible population. Bullets mark cells that are identically zero.

|         | IDU   | MSM   | FSW   | MCF   | FPM   | MSM &IDU | FSW &IDU |
|---------|-------|-------|-------|-------|-------|----------|----------|
| IDU     | 19.27 | •     | •     | •     | •     | 19.27    | 19.27    |
| MSM     | •     | 4.09  | •     | •     | •     | 4.09     | •        |
| FSW     | •     | •     | •     | 0.058 | •     | 0.000    | •        |
| MCF     | •     | •     | 77.27 | •     | •     | •        | 77.27    |
| FPM     | 0.007 | 0.001 | •     | 0.010 | •     | 0.000    | •        |
| MSM&IDU | 2.001 | 1.18  | •     | •     | •     | 3.329    | 2.00     |
| FSW&IDU | 0.469 | •     | •     | 0.003 | •     | 0.469    | 0.93     |
| Total   | 22.00 | 5.27  | 77.27 | 0.071 | •     | 27.150   | 99.47    |
| Eigenvector | 0.989 | 0.025 | 0.000 | 0.088 | 0.000 | 0.111 | 0.025 |

If all of the connections between groups were broken, the IDU epidemic would still be self-sustaining ($R_0 = 19$). Similarly, the epidemic in FSW and MCF and the epidemic in MSM would both be self-sustaining but transmission in these groups is much easier to control because $R_0$ is smaller ($R_0 = \sqrt{77.27 \times 0.058} = 2.1$ and 4.1, respectively). Because the IDU epidemic is linked to both MSM and FSW, control in the whole network will ultimately depend on control in IDUs.

Considering the links between pairs of groups (Figure 3), the most important are between IDU and MSM who are also IDU ($R_0 = \sqrt{19.27 \times 2.00} = 6.2$) or FSW who are also IDU ($R_0 = \sqrt{19.27 \times 0.469} = 3.0$). The loop connecting FSW and MCF is highly asymmetric: one case introduced in the FSW population will infect 77 MCF on average but each MCF infects only 0.058 FSW on average, over the life-time of an infected person. Thus the number of secondary cases arising in FSW via MCF, over one complete cycle of transmission, is $\sqrt{77 \times 0.058} = 2.1$. There are considerably fewer HIV-positive FSW than MCF (81 *versus* 653, Table 1) and each FSW has the potential to infect many more MCF over the ten years for which they will survive without treatment (77 *versus* 0.058, Table 2) so that, per person treated, interventions aimed at stopping transmission to and from FSW will be much more effective than interventions aimed at MCF.

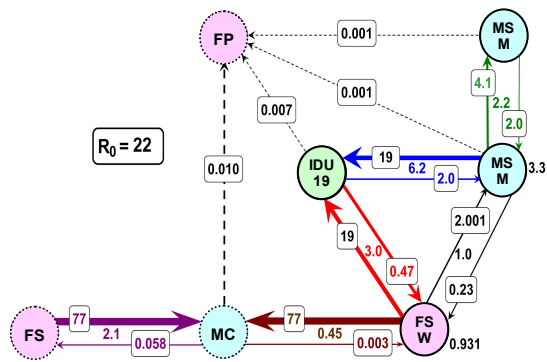

Figure 3. The network model with estimated transmission rates, which form the components of the next-generation matrix. Each number is the total number of secondary infections arising from one infected case in an otherwise susceptible population. Numbers in circles are for transmission within a population sub-group; numbers on lines are for transmission from one group to another; numbers between lines are the number of infections transmitted around a loop.

The epidemic of HIV in MSM ($R_0 = 2.2$) should also be relatively easy to control. The prevention and treatment of infection in FPM is important in its own right and infected women should all have access to life-saving anti-retroviral therapy, but these women are assumed to be an epidemiological dead end, so control measures will not affect transmission elsewhere in the network.

To choose the most effective control measures, the values of $R_0$ need to be considered in relation to the efficacy of different interventions. To eliminate the epidemic within the IDU population requires $R_0$ to be reduced by a factor of more than 19 (i.e. by more than 95%). Even with widespread use of antiretroviral therapy (ART)—for which the efficacy in preventing transmission from HIV-positive people to others has been estimated at 96% [12]—this is a challenge for ART when used as a single intervention and would demand very high levels of compliance and viral load suppression. However, combining ART with an effective needle exchange programme (a new needle carries zero risk of acquiring HIV) and opiate substitution therapy to reduce the use of injectable drugs, should be sufficient to achieve $R_0 < 1$ in the IDU population.[16] If, for example, half of the needle sharing involves clean needles then this would reduce $R_0$ to about 10 and one would then only need a further 90% reduction through the use of ART to reduce $R_0$ to less than 1; if clean needles were used in 95% of risky injection events then this alone would reduce $R_0$ below 1 without the need for ART.

For each of the FSW and MSM populations it would be necessary to reduce transmission by a little more than 50% (more than 55% for $R_0 = 2.2$ in MSM). For FSW, condom promotion would have a major impact and a '100% condom programme' of the kind carried out in Thailand[17] should be sufficient to bring $R_0$ below 1 for FSW and MCF, especially if supported by universal access to ART.[11] While consistent and correct condom use will completely stop the transmission of HIV, MSM may be reluctant to use condoms and condom promotion is generally found to be much less effective even under trial conditions.[18] However, a programme of condom promotion combined with regular testing and universal access to ART should reduce $R_0$ by more than the factor of 4.1 needed to control the epidemic among MSM.

Elimination of HIV from the whole network requires a combination of interventions against IDU, MSM, FSW and other groups. Further insights into the best combination of interventions that most effectively reduce $R_0$ are provided by the NGM. Formally, elimination requires not only that $R_0 < 1$ for each population group, but also for the network overall. Focusing on the key groups of IDU, MSM and FSW, let us assume that different numbers of each can be removed from the pool of potentially infectious people, either by ensuring that viral load is fully suppressed for those already infected, or that uninfected individuals can be fully protected against infection through the use of clean needles, opiate substitution therapy, of condom promotion. For any combination of IDU, MSM and FSW that is rendered non-infectious in this way we calculate the resulting value of $R_0$ for the whole network. In what follows use the word 'treatment' to indicate that people have been rendered non-infectious.

Figure 4A and 4B show, in principle, how to minimize $R_0$ for the whole network by making the smallest number of people non-infectious. Notice that in each panel the lower contours of constant $R_0$ are almost flat, so the best way to reduce $R_0$ initially is by treating IDU alone (Figure 4A, vertical axis); there is little to be gained by treating members of other groups until a sufficiently large number of IDU have effectively been removed from the transmission network. The yellow dot in Figure 4A marks the position at which 2,400 IDU, but no MSM, are treated. Then, moving from the yellow dot, the best strategy is to treat both IDU and MSM until 2,700 IDU and 900 MSM are on treatment as indicated by the blue dot in Figure 4A. If one were only going to intervene with IDUs and MSM one would then continue along the line to the top right hand corner of Figure 4A when 3,100 IDUs and 1,150 MSM were on treatment. However, this would not eliminate transmission from the network as the epidemic in FSWs and MCFs is self-sustaining and the value of $R_0$ for the whole network would be 2.2. The optimal strategy, after reaching the light blue point in Figure 4A or the corresponding light blue point in Figure 4B would be to increase the number of IDUs and MSM on treatment, keeping the proportion of each constant, but start treating FSWs following the curved line to the dark blue point in Figure 4B when $R_0$ for the whole network would be reduced to 1. After that one would continue to the red dot in Figure 4B when all IDU, MSM and FSWs are rendered uninfectious and $R_0$ for the whole network is reduced to 0.12.

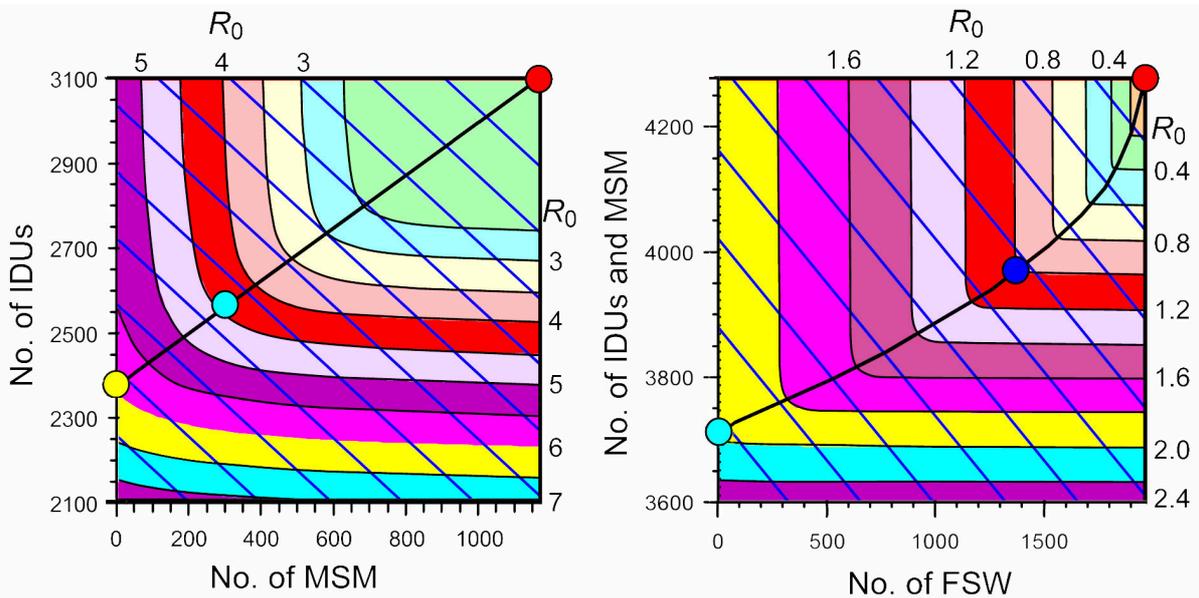

Figure 4. Surface plots of $R_0$ as a function of the number of IDUs, MSM and FSWs who are rendered uninfectious either through treatment or prevention interventions. A: number of IDU plotted against the number of MSM who are rendered uninfectious. B: combined number of IDU and MSM plotted against the number of FSW who are rendered uninfectious. Shaded areas give contours of constant $R_0$ for the values shown above and to the right of each plot. Diagonal lines in A indicate combinations of MSM and IDUs for fixed total numbers from 2,100 (bottom left) to 4,278 (top right); in B they indicate fixed total numbers of MSM and IDU (vertical axis) and FSW (horizontal axis) from 3,600 (bottom left) to 6,200 (top right). The lines running across each plot indicate the optimal combinations of IDUs, MSM and FSWs that minimize $R_0$ for A: a fixed total number of IDUs and MSM and B: a fixed total number of IDUs, MSM and FSW.

The virtue of the NGM is that it gives an instant analytical guide to the question of where to focus interventions. To confirm the above results and also to explore the impact of different interventions through time demands a full dynamical simulation and this is illustrated in Figure 5.

Figure 5A shows the expected prevalence and incidence of HIV and AIDS-related mortality without any intervention in any group. This corresponds to the model fits given in Figure 2, projected forwards to 2050. Figure 5B shows what would happen in all five population groups if all IDU, but only IDU, were treated within one year of acquiring HIV infection so as to eliminate onward transmission. In Figure 5C to 5F the calculation is repeated for FSW, MCF, MSM and FPM.

Naturally, the treatment of people in any group reduces incidence, prevalence and mortality in that group. However, a comparison of Figure 5B with Figure 5C to 5F shows that only the treatment of IDU has a major impact on infections in all other population groups in the network. The secondary effect on MSM is most rapid, followed by FSW, MCF and FPM, as expected from the network structure shown in Figure 3. The treatment of FSW is also very beneficial for MCF (Figure 5C), but the reverse is not true (Figure 5D). Infection cannot be eliminated from the network by treating any one population group alone, though the treatment of IDU has the biggest overall impact. The benefits for other groups of treating MSM are small because MSM are weakly linked in the network (Figure 5E), except for the small proportion that also inject drugs (Figure 3). There are no benefits for other groups of treating FPM, because they are assumed not to transmit infection to anyone else (Figure 5F).

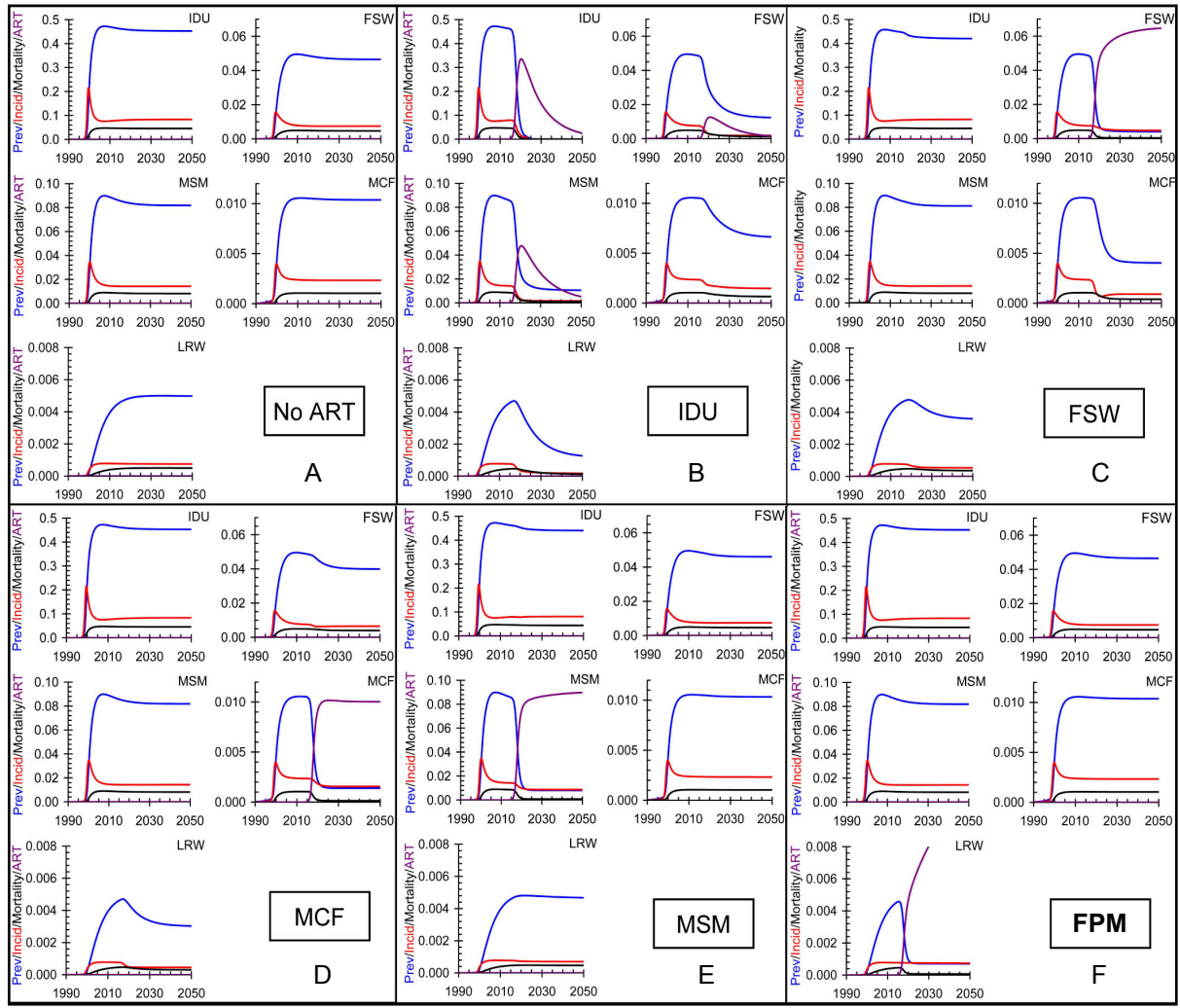

Figure 5. The boxes show the projected prevalence, incidence, mortality and ART coverage assuming that ART is provided to all those in the relevant population, as indicated in each box, with 50% coverage being reached in 2015 and full coverage in 2020 (see text for details). Blue lines: prevalence, red lines: annual incidence; black lines: annual mortality; purple lines: prevalence of people on ART.

## 4. Discussion

The better we understand the structure of transmission networks, the more effectively we can target efforts in infectious disease control. For concentrated epidemics of HIV/AIDS, the assumption of random mixing greatly underestimates the contribution of some population groups and greatly overestimates the contribution of others. This is illustrated here by the large difference in estimates of the basic case reproduction number of HIV estimated by assuming homogenous mixing ($R_0 = 4$; data not shown) and derived from the structured network that describes ($R_0 = 22$).

There is a cost to investigating the detailed structure of transmission networks, but the approach suggested here requires only data that are routinely collected during the spread of an epidemic, disaggregated for population groups that are likely to be exposed to infection at different rates, or transmit infection by different routes. As a further demonstration of heterogeneity, our reconstruction of the HIV/AIDS epidemic in Can Tho province shows that the infection was probably introduced first in FSW and MCF but then spread to IDU which became the main drivers of the epidemic. Ultimately the network generated most cases among the female partners of sex worker clients.[14]

The structure and value of the elements in the next generation matrix give a guide to the key points at which control must be implemented and the degree of control that is needed to bring the epidemic to an end. Our analysis shows that IDU are the largest contributors ($R_0 = 19.3$) to the overall case reproduction number ($R_0 = 22.0$). The control of infection in IDU is the most effective way to reduce infections, not only in IDU, but across the whole network. To eliminate infection from the network altogether requires the reduction of $R_0$ in IDU by a factor $\geq 19$, but this must be achieved in combination with treatments for other population groups, especially FSW and MSM, so that $R_0 < 1$ for every group separately and $R_0 < 1$ for the network overall. Numerical simulations confirm these results, and show in detail how HIV incidence and prevalence and AIDS-related mortality can be expected to change through time.

The optimal combination of prevention methods will depend on the group being targeted. For IDUs one would need a combination of opiate substitution therapy or methadone maintenance,[11] access to clean needles,[10] social support and ART[12] as soon as people are found to be living with HIV. By combining these interventions one would get significant synergies. If, as is the case in Vietnam, methadone maintenance programmes require daily attendance by patients with a medical doctor present at all times, this would provide an ideal setting for the provision of anti-retroviral drugs combined with testing viral loads to ensure compliance. For FSWs in brothels a condom programme of the kind rolled out in Thailand should have a significant effect on transmission[9] and this could be combined with routine testing of women for HIV, if they are previously HIV-negative, and viral loads, if they are already infected with HIV.

Although the NGM constructed from routine data gives insights into HIV epidemiology and control quickly and relatively easily, it is not the last word in analysis. For instance, genotyping studies could help to confirm or refute our deduction that the epidemic in IDU was first introduced by FSW. We have also assumed that the population of Can Tho province is affected by a single strain of HIV even though, in other settings, MSM may be infected by different strains of HIV from FSW and MCF giving rise to separate epidemics.[19] Because our method of analysing epidemic spread and control requires only routine data, it could potentially be applied to HIV infection and other communicable diseases in different populations. However, it would be instructive and prudent to carry out an analysis of HIV strain variation in HIV infections in any other population to which this method of network analysis might be applied.

Here we are concerned to demonstrate the information that can be gained from a detailed analysis of the structure of the NGM as a guide to the choice of interventions and to facilitate a complete analysis of future projections, estimates of impact, and costs and cost effectiveness of different interventions. In this, as in all public health data, there is uncertainty in the data, and hence in the fitted curves and corresponding parameter estimates and these should be used to add uncertainty estimates to the various fitted parameters, estimates of $R_0$ and future projections. It is, of course, important to bear in mind that many different objective functions can be chosen; here we have chosen to minimize $R_0$, others might wish to minimize the total cost of the intervention, the cost per infection averted or per life saved, for example, and each of these would lead to a different optimal strategy. Furthermore, in each particular setting it will be necessary to identify the relevant risk groups, decide on the optimal combination of interventions for each group taking into account both the efficacy and the cost of each component intervention, and then plan the roll-out of the control programme accordingly. Finally, a full dynamical model should be used to evaluate the long term impact, on the HIV prevalence, incidence and mortality as well as the cost and cost-effectiveness of alternative interventions An analysis of the kind presented in this paper should, however, provide a useful and informative starting point for thinking about the best way to control HIV.

## Appendix: Model equations

The model used in this analysis is illustrated schematically in Figure 1. The structure, and in particular, the overlapping groups and the links between pairs of groups, was arrived at after extensive discussions with field workers supporting each of the risk groups in Vietnam.[14] A critical risk group consists of female sex workers who also inject drugs and form a bridging population between those who are at risk through heterosexual transmission and those who are at risk through the sharing of contaminated needles. Most of these women are primarily injecting drug users who do sex work to raise money to buy drugs. The equations for the model in Figure 3 are:

$$S_d^\bullet = (\mu_d + \delta)I_d - \beta_d P_d S_d \qquad 1$$

$$I_d^\bullet = \beta_d P_d S_d - (\mu_d + \delta)I_d \qquad 2$$

$$S_m^\bullet = (\mu_m + \delta)I_m - \beta_m P_m S_m \qquad 3$$

$$I_m^\bullet = \beta_m P_m S_m - (\mu_m + \delta)I_m \qquad 4$$

$$S_s^\bullet = (\mu_s + \delta_s)I_s - \beta_s P_c S_s \qquad 5$$

$$I_s^\bullet = \beta_s P_c S_s - (\mu_s + \delta)I_s \qquad 6$$

$$S_{md}^\bullet = (\mu_{md} + \delta)I_{md} - (\beta_m P_m + \beta_d P_d) \qquad 7$$

$$I_{md}^\bullet = \beta_m P_m + \beta_d P_d - (\mu_{md} + \delta)I_{md} \qquad 8$$

$$S_{sd}^\bullet = (\mu_{sd} + \delta)I_{sd} - (\beta_s P_c + \beta_d P_d) \qquad 9$$

$$I_{sd}^\bullet = \beta_d P_d + \beta_s P_c - (\mu_{sd} + \delta)I_{sd} \qquad 10$$

$$S_c^\bullet = (\mu_c + \delta)I_c - (\beta_s P_s + \mu_c)S_c \qquad 11$$

$$I_c^\bullet = \beta_c P_s S_c - (\mu_c + \delta)I_c \qquad 12$$

$$S_w^\bullet = (\mu_w + \delta)I_w - (\beta_w I_c + \mu_w)S_w \qquad 13$$

$$I_w^\bullet = \beta_w I_c S_w - (\mu_w + \delta)I_w \qquad 14$$

In Equations 1 to 14 $S$ refers to susceptible people, $I$ to infected people, $\mu$ determines the rate at which people leave each of the groups so that $1/\mu$ is the mean time for which a person remains in that group in the absence of HIV and $\beta$ gives the rate of transmission for people in each group. The subscripts $d$ refer to IDU only, $m$ to MSM only, $s$ to FSW only, $c$ to MCF, $w$ to FPM, $md$ to MSM who are also IDU, $sd$ to FSW who are also IDU. $P_i$ is the average prevalence among people in group $i$ so that $P_d$, for example, is the prevalence of HIV averaged over all those that use drugs including those who only use drugs as well as FSW and MSM who use drugs. $\delta$ without a subscript refers to the background mortality which we take to be the same for all groups. The chance of being infected through drug use is independent of whether or not that person is also MSM or FSW. In practice MSM who also use drugs may be more likely to be infected by other MSM rather than FSW who also use drugs

In order to allow for heterogeneity in risk, which determines the steady state prevalence of infection, we assume that the rates of transmission ($\beta$ in these equations) are multiplied by a corresponding Gaussian term so that

$$\beta_i = \beta_i^0 \, e^{-\alpha_i P_i^2} \qquad 15$$

so that $\beta_i^0$ is the rate of transmission in group $i$ at the start of the epidemic when the prevalence is close to zero and the rate of transmission fall as the prevalence $P_i = I_i/N_i$ rises. At the start of the epidemic the prevalence is low but those at highest risk will be infected first. As prevalence rises, those that are not yet infected will be at lower risk and the average value of the transmission parameter will decrease as the prevalence of infection increases. Previous studies have assumed an exponential relationship or a step-function. In the former case the solutions tend to be unstable as the risk of infection drops rapidly and prevalence increases rapidly as prevalence declines. In the latter case one is dividing the population into those at a certain fixed risk and those at no risk. The prevalence data can be fitted equally well under either of these two extreme assumptions and there is, unfortunately, no direct evidence to determine the rate at which transmission falls as prevalence rises. This is an area that warrants further investigation.[1]

The model fits are given in Figure 2 and the parameter values for the fits are given in Table A1. The time for which people remain in a risk group and the size of each risk group were obtained from field workers supporting each of the risk groups in Vietnam.[14]

Table A1. Best fit values of the transmission parameters, estimated durations within each risk group and estimated size of each risk group. The loss rate is the rate at which people leave each group. The AIDS related mortality is 0.1/year.

| Transmission/yr | | Loss rate/yr | | Group size | |
|---|---|---|---|---|---|
| $\beta_d$ | 2.51 | $\mu_d$ | 0.084 | $N_d$ | 3,698 |
| $\beta_m$ | 0.08 | $\mu_m$ | 0.050 | $N_m$ | 1,183 |
| $\beta_c$ | 0.34 | $\mu_c$ | 0.100 | $N_s$ | 2,011 |
| $\beta_s$ | 0.24 | $\mu_s$ | 0.125 | $N_{md}$ | 384 |
| $\beta_w$ | 0.00 | $\mu_w$ | 0.050 | $N_{sd}$ | 90 |
| | | | | $N_c$ | 62k |
| | | | | $N_w$ | 455k |